# An Identity and Interaction-Based Network Forensic Analysis

Nathan Clarke
University of Plymouth, Plymouth, UK
nathan.clarke@plymouth.ac.uk

Gaseb Alotibi
University of Tabuk,
galotaibi@ut.edu.sa

Dany Joy
University of Plymouth, Plymouth, UK
dany.joy@plymouth.ac.uk

Fudong Li
Bournemouth University
fli@bournemouth.ac.uk

Steven Furnell
University of Nottingham
steven.furnell@nottingham.ac.uk

Ali Alshumrani
University of Plymouth, Plymouth, UK
ali.alshumrani@plymouth.ac.uk

Hussam Mohammed
University of Anbar
hussamjasim@uoanbar.edu.iq

Prof. Nathan Clarke is a Professor in Cyber Security and Digital Forensics at the University of Plymouth, UK. He is also an adjunct Professor at Edith Cowan University in Australia. His research interests reside in cyber security, specifically biometrics, digital forensics and artificial intelligence. Prof Clarke has published over 250 journal and conference papers. He co-created and co-chairs the *International Symposium on Human Aspects of Information Security and Assurance (HAISA)*, currently in its nineteenth year. Prof Clarke has been involved in a number of successful EPSRC, Knowledge Transfer Projects and European funded projects and has graduated 48 doctoral students. Prof Clarke is a chartered engineer, a fellow of the British Computing Society (BCS) and a senior member of the IEEE.

Dr Gaseb Alotibi received the B.Sc. degree in information system from King Saud University, S.A, and the M.Sc. degree in computer system security from South Wales University and the Ph.D. degree in Cybersecurity from University of Plymouth, UK. His Ph.D. research focused on behavioral profiling via network communication. He is currently working as an Assistant Professor at University of Tabuk. His current research interests include digital forensics, digital health and cybersecurity.

Dr Dany Joy received his PhD in 2001 from the University of Plymouth. His work was focused upon the development of innovative approaches to the forensic analysis of network data.

Dr Fudong Li is a reliable and experienced Principal Academic in Cyber Security at Bournemouth University UK. Dr Li has a broad range of research interests, including biometric authentication, digital forensics and network security. He has 12 PhD completions and published over 60 journal and international conference papers. In addition, he has several industrial certifications, including EC-Council certified Computer Hacking Forensic Investigator and CCNA (ITN).

Prof. Steven Furnell is Professor of Cyber Security in the School of Computer Science at the University of Nottingham. His research interests include security management and culture, usability of security and privacy, and technologies for user authentication and intrusion detection. He has authored over 400 papers in refereed international journals and conference proceedings, as well as various books, book chapters, and industry reports. Steve is the UK representative to Technical Committee 11 (security and privacy) within the International Federation for Information Processing, and a board member of the Chartered Institute of Information Security, and a member of the Steering Group for the Cyber Security Body of Knowledge (CyBOK) and the Careers and Learning Working Group within the UK Cyber Security Council.

Ali Alshumrani is a PhD candidate specialising in Digital Forensics at the Centre for Cyber Security, Communications and Network Research (CSCAN) at the University of Plymouth. His research focuses on developing a unified digital forensics tool to improve investigative processes, enhance data analytics automation, and reduce the cognitive load on forensic investigators, ultimately enabling more efficient and effective cybercrime investigations. His areas of interest include digital forensics, cyber security, data ontology, and knowledge graphs.

Dr Hussam J. Mohammed is a Doctor of Philosophy specializing in Digital Forensics and Artificial Intelligence. He earned his BSc and MSc degrees in computer science from the University of Anbar, Iraq, before pursuing his PhD in digital forensics and cybersecurity at the University of Plymouth, UK. Currently, Hussam is a faculty member in the AI Department at the University of Anbar, where he is actively involved in research and academic development. His primary research interests include digital forensics, machine learning, artificial intelligence, and the application of AI in cybersecurity and digital evidence. Hussam strives to bridge the gap between theoretical research and practical applications, contributing significantly to the future of AI-driven security technologies

# An Identity and Interaction-Based Network Forensic Analysis

*Nathan Clarke*
*University of Plymouth*
*nathan.clarke@plymouth.ac.uk*
*Dany Joy*
*University of Plymouth*
*dany.joy@plymouth.ac.uk*

*Gaseb Alotibi*
*University of Tabuk*
*galotaibi@ut.edu.sa*
*Fudong Li*
*Bournemouth University*
*fli@bournemouth.ac.uk*

*Steven Furnell*
*University of Nottingham*
*steven.furnell@nottingham.ac.uk*

*Ali Alshumrani*
*University of Plymouth*
*ali.alshumrani@plymouth.ac.uk*

*Hussam Mohammed*
*University of Anbar*
*hussamjasim@uoanbar.edu.iq*

## Abstract

In today's landscape of increasing electronic crime, network forensics plays a pivotal role in digital investigations. It aids in understanding which systems to analyse and as a supplement to support evidence found through more traditional computer-based investigations. However, the nature and functionality of the existing Network Forensic Analysis Tools (N-FATs) fall short compared to File System Forensic Analysis Tools (FS-FATs) in providing usable data. Current N-FATs often present data at an overly granular level, making it challenging for investigators to extract meaningful insights in a timely manner. Moreover, the analysis tends to focus on IP addresses, which are not synonymous with user identities, a point of significant interest to investigators. This paper presents several experiments designed to create a novel N-FAT approach that can identify users and understand how they are using network-based applications whilst the traffic remains encrypted. The experiments build upon the prior art and investigate how effective this approach is in classifying users and their actions. Utilising an in-house dataset composed of 50 million packets, the experiments three incremental developments that assist in improving the performance. Building upon the successful experiments, a proposed N-FAT interface is presented to illustrate the ease at which investigators would be able ask relevant questions of user interactions. The experiments profiled across 27 users, has yielded an average 93.3% True Positive Identification Rate (TPIR), with 41% of users experiencing 100% TPIR. Skype, Wikipedia and Hotmail services achieved a notably high level of recognition performance. The study has developed and evaluated an approach to analyse encrypted network traffic more effectively through the modelling of network traffic and to subsequently visualise these interactions through a novel network forensic analysis tool.

# An Identity and Interaction-Based Network Forensic Analysis

## 1 Introduction

Digital forensics has become pivotal in investigating cyber and computer-assisted crimes, with a historical focus on computer systems and File-System Forensic Analysis Tools (FS-FATs) and their accompanying application-level parsers. However, the recent surge in smartphone popularity has also led to the prominence of mobile adaptations of these tools. While these solutions have demonstrated success, evolving technology, the dynamic threat landscape, and the emergence of anti-forensic tactics have underscored the increasing significance of network forensics, which stands as an independent and autonomous source of evidence beyond the reach of adversaries (Hasanabadi et al. 2020).

Existing Network Forensic Analysis Tools (N-FATs) like Wireshark and Xplico have primarily served as network protocol analysers for administrators, offering limited forensic capabilities (Khan et al. 2016). These tools operate at a low network packet level, hindering investigators' ability to ask high-level questions in a cognitively simple and timely manner due to the sheer volume of network data, making sifting through packets time-consuming (Alotibi et al. 2016). Furthermore, these tools often fail to manage and handle the data in a manner that investigators would expect.

Therefore, this paper proposes an N-FAT tailored for investigations focusing on suspects, addressing key questions about their activities, interactions, and associates. This paper leverages user interactions with network-based applications to identify users and provide high-level network data for rapid, meaningful analysis. The main contribution of the N-FAT includes a series of experiments that are designed to enable investigators to search encrypted network traffic based on users and applications rather than Internet Protocol (IP) addresses. Having established the feasibility of the approach, the study discusses how this understanding of network traffic can then be utilised by investigators to more easily understand what has happened, by whom and when.

The subsequent sections of the paper are structured as follows: Section 2 presents the related literature, emphasising limitations within existing approaches. Section 3 presents the experimental methodology, while Section 4 provides the experimental results. Section 5 discusses the utility of the approach and how this can be incorporated into an N-FAT. Finally, Section 6 concludes the paper and outlines future research directions.

## 2 Analysis of the Prior Art

This section will include a review aimed at conducting an analysis of the prior art. It will serve as a foundation for comprehending the concepts necessary to appreciate the novel contribution of this research.

### 2.1 Packet-based and Flow-based Methods

Various methods have been developed to detect, monitor, understand, or prevent network-related incidents and attacks. These approaches primarily operate using two methods for examining network data: packet-based analysis, also known as Deep Packet Inspection (DPI), which examines the contents of IP packets to identify data and detect threats, and flow-based analysis, which utilises IP flows to summarise related packets with shared properties, including timestamps, IP addresses, port numbers, packet count, size, and traffic type.

Table 1 presents an overview of various N-FATs, delineating their distinct roles in the surveillance and examination of network traffic. Predominantly, these tools excel in capturing, monitoring, reconstructing, detecting, and analysing network-related incidents. Notably, tools like Wireshark and TCPdump possess the capability to decrypt traffic when encryption keys are accessible (Tcpdump, 2023; Wireshark, 2023). However, it is imperative to acknowledge that these tools are not inherently designed to breach encryption or autonomously profile network traffic content. Furthermore, certain tools lack advanced dashboard functionality, constraining investigators from implementing specific filters such as timestamps, protocols, and IP addresses to generate comprehensive summary information. In contrast, the proposed system is centred on identifying users and deciphering their utilisation of network-based applications whilst maintaining the encrypted state of the traffic. This involves a meticulous examination of metadata, packet sizes, timing, and endpoints, circumventing the need to access the actual content of the packets.

**Table 1: Existing Network Forensics**

| Tool Name | Analytical Approach | License | Graphical Interface | Main Feature(s) | Decryption Capabilities |
|---|---|---|---|---|---|
| Nagios | Flow | Free | Yes | Monitoring and alerting system | Incapable |
| NetworkMiner | Packet | Required | Yes | Examine, reconstruct, and visualise network sessions | Incapable |
| OpenNMS | Flow | Free | Yes | Network performance monitoring | Incapable |
| Pandora FMS | Flow | Free | Yes | Comprehensive monitoring solution | Incapable |
| Pyflag | Packet | Free | Yes | Network traffic analysis | Limited capabilities |
| Splunk | Flow | Required | Yes | Data collection and analysis | Incapable |
| Tcpdump | Packet | Free | No | Network traffic capture and analysis | Requires keys |
| Wi-Fi Network Monitor | Flow | Free | Yes | Wireless network monitoring | Incapable |
| WirelessNetView | Flow | Free | Yes | Wireless network monitoring | Incapable |
| Wireshark | Packet | Free | Yes | Network traffic capture and analysis | Requires keys |
| Xplico | Flow | Free | Yes | Internet traffic extraction and reconstruction | Requires keys |

**Table 2: Examples of Existing Network Monitoring Studies**

| Reference | Approach | Applications | Performance |
|---|---|---|---|
| (Ahmed and Lhee, 2011) | Packet-based | Malware Detection | 4.69% false negative rate, 2.53% false positive rate |
| (Al-Bataineh and White, 2012) | Packet-based | Data Exfiltration Detection | 99.97% detection rate on HTTP traffic |
| (He et al. 2014) | Packet-based | Data Exfiltration Detection | 90% detection rate, less than 1% false positives |
| (Parvat and Chandra, 2015) | Packet-based | IDS | 98.5% correct classification rate |
| (Boukhtouta et al. 2016) | Packet-based | Malware Classification | 99% precision, less than 1% false positives |
| (Stergiopoulos et al. 2018) | Packet-based | Malicious Traffic Detection | 94% true positive detection rate |

| (Tegeler et al. 2012) | Flow-based | Malware Detection | 90% detection rate, 0.1% false positive rate |
|---|---|---|---|
| (Hofstede et al. 2013) | Flow-based | Anomaly-based Network IDS | 95% detection rate, 1% false positive rate |
| (Stevanovic and Pedersen, 2014) | Flow-based | Anomaly Detection | Numerical values are not provided |
| (Fernandes et al. 2015) | Flow-based | IDS | 99.4% detection rate, 0.6% false alarm rate |
| (Taylor et al. 2016) | Flow-based | Applications Identification | 99% accuracy rate in re-identifying profiled apps |
| (Clarke et al. 2017) | Flow-based | User Identification | Up to 90% recognition rates |
| (Leroux et al. 2018) | Hybrid | Traffic Classification | Numerical values are not provided |
| (Meghdouri et al. 2020) | Flow-based | Anomaly Detection | Numerical values are not provided |

The studies in Table 2 showcase a diverse array of methodologies for analysing and securing network traffic, predominantly concentrating on malware detection, classification, data exfiltration detection, IDS, anomaly detection, traffic classification, botnet detection, and application identification. The table also highlights the proficient performance of both DPI-based and flow-based methods in detecting and classifying a wide spectrum of network events. Although these studies employ diverse methodologies and pursue distinct objectives, they collectively emphasise the necessity for advanced approaches to network traffic analysis, especially in response to the growing prevalence of encryption. This body of research highlights a notable trend in network forensics and traffic analysis. Studies such as those by (Ahmed and Lhee, 2011; Al-Bataineh and White, 2012; He et al. 2014) delve into the complexities of analysing network payloads and encrypted traffic, elucidating the challenges of accurately identifying and categorising data amidst encryption. The necessity for sophisticated mechanisms to discern between different types of content and the importance of statistical features and behaviour profiling in encrypted environments are consistent threads.

The works of (Boukhtouta et al. 2016; Parvat and Chandra, 2015; Stergiopoulos et al. 2018; Stevanovic and Pedersen, 2014; Tegeler et al. 2012) further reinforce the potential of integrating machine learning techniques and heuristic-based methods in network traffic analysis. Collectively, these studies demonstrate high precision in detecting and classifying network activities, emphasising the effectiveness of data-driven approaches. Moreover, the integration of IDS directly into network infrastructure, as explored by (Hofstede et al. 2013), and the employment of neural networks and Principal Component Analysis (PCA) for anomaly detection and traffic profiling, as seen in (Abuadlla et al. 2014; Stevanovic and Pedersen, 2014) represent significant strides towards real-time, accurate network monitoring and threat mitigation.

Despite the robust methodologies and significant detection accuracies presented in these studies, a common limitation is their reliance on decryption or superficial analysis when dealing with encrypted traffic. This constraint often leads to a trade-off between user privacy and analytical depth. The reviewed literature primarily offers insights into the type and nature of network traffic, with less emphasis on understanding user behaviour and application usage patterns in an encrypted environment. This gap underscores the necessity for an advanced N-FAT approach that can delve deeper into encrypted traffic, providing comprehensive insights without compromising encryption integrity. In light of these findings, the proposed N-FAT approach in this study aims to fill a critical gap in the current landscape of network forensics. While the reviewed literature and N-FATs predominantly provide low-level information, the N-FAT approach seeks to transcend these boundaries. It aims not only to identify users but also to understand their behaviour in network-based applications without decrypting the traffic. This contribution is poised to address a pivotal need in network forensics, offering a nuanced, comprehensive tool for network traffic analysis that respects the integrity of encryption while providing deep insights into user behaviour and network usage.

## 2.2 Interactions and Behavioural Profiling

Compared to the mentioned techniques, (Clarke et al. 2017) introduced a study demonstrating the use of unique user interactions at the network level. These interactions allow the identification of individual actions users perform on network-based applications, even with encrypted traffic, without decryption or DPI. The study involved examining network traces generated during interactions, enabling the identification of specific user actions rather than just network signals. A dataset from 46 users over 60 days was used, containing metadata like timestamps, IP addresses, port numbers, packet length, traffic type, and flags. The approach employed a single Feed-Forward Multi-Layer Perceptron (FF-MLP) neural network in identification mode, achieving a promising 90% recognition rate. While promising, a single classifier used in identification mode will likely struggle to scale appropriately and potentially require a large complex neural network with a subsequent computational impact. To this end, this paper presents a series of further experiments to develop the N-FAT approach as an enabling platform to aid the forensic investigation of network data.

## 3 Identity and Service-Based Detection: Experimental Methodology

A key contribution of this research is the ability to attribute interactions to individuals rather than to IP addresses. The evaluation aims to determine if these methods improve recognition levels compared to using a single classifier as in (Clarke et al. 2017). While the single classifier approach performed well, it would face scalability challenges with a growing user population. As such, an alternative strategy explored was the use of $n$ 2-class classifiers addressing scalability and potentially providing better recognition granularity. This formed the basis of the first experiment. Research in multibiometrics also indicates performance improvements through fusion, particularly classification-level fusion (Saevanee et al. 2015), as shown in Figure 1 (and this formed the basis of the second experiment).

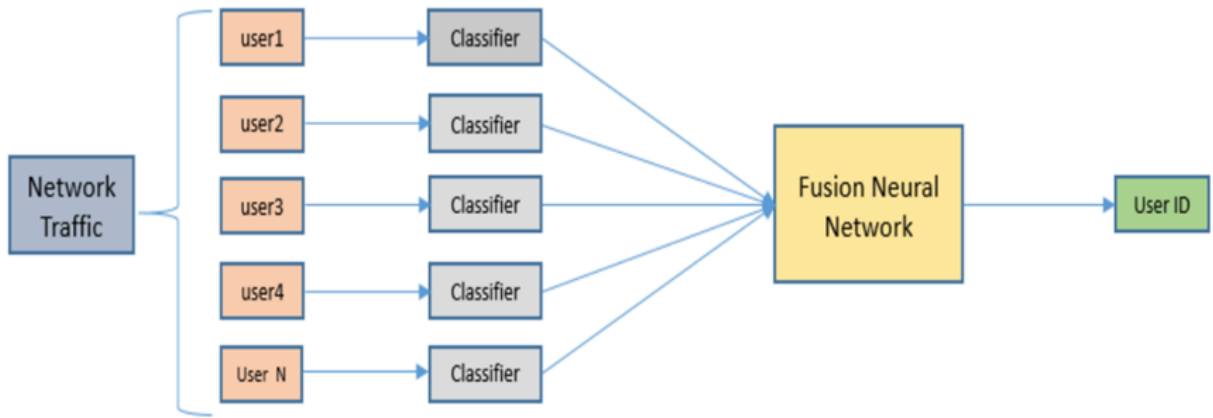

**Figure 1: A Fusion-based Approach to User Identification**

Whilst IP addresses are not static due to Dynamic Host Configuration Protocol (DHCP) and ad hoc mobile devices connecting and leaving networks, within a short time frame they can be assumed to be static. The third experiment sought to investigate the impact upon performance through applying this assumption of IP addresses over short time windows (seconds to a few minutes). If the classifier confidently identifies a user from a sample, it can infer that all IP traffic within a certain time window before and after that sample is also from the same user. This algorithmic approach helps mitigate weak classification decisions. To maintain consistency with prior research and due to the absence of more suitable datasets, an in-house dataset was employed (Alotibi et al. 2016). This dataset comprised data collected from 27 users over two months. During this period, participants were asked to use their computers normally, aiming to capture authentic user behaviour. All network traffic was monitored, and IP header metadata was recorded. To calculate error rates, users' source IP addresses remained constant. Participants were instructed not to share their systems during data collection to ensure the legitimate user's application behaviours were captured. The dataset contains IP header information from over 140 million packets.

**Table 3: Experimental Dataset: Service Overview**

| Application | Number of packets | Number of Interactions | Data Reduction % | Number of Participants |
|---|---|---|---|---|
| YouTube | 21,131,316 | 1,322,848 | 93.8 | 27 |
| Facebook | 5,727,953 | 386,741 | 93.3 | 27 |
| Google | 1,857,420 | 194,404 | 89.6 | 27 |
| Twitter | 747,584 | 71,403 | 90.5 | 27 |
| Wikipedia | 1,250,302 | 5,719 | 99.5 | 20 |
| Hotmail | 703,711 | 122,989 | 82.6 | 19 |
| Dropbox | 17,480,739 | 98,555 | 99.4 | 16 |
| BBC | 201,263 | 4,180 | 97.9 | 12 |
| Skype | 575,030 | 178,686 | 68.9 | 12 |

As in the previous study (Clarke et al. 2017), application-level interactions were identified among the monitored nine services for which interaction signatures had been previously identified. Table 3 provides an overview of the dataset by service across the population, while Table 4 breaks it down by user. Table 3 also illustrates the data reduction process, reducing interactions from 50 million raw packets to 2.4 million interactions. This significant reduction reduces the cognitive load for machine learning and investigators. Notably, not all users used all nine applications, resulting in varying participation per application.

**Table 4: Experimental Dataset: Individual User and Service**

| User ID | BBC | Dropbox | Facebook | Google | Hotmail | Skype | Twitter | Wikipedia | YouTube | Total |
|---|---|---|---|---|---|---|---|---|---|---|
| 1 | 0 | 0 | 528 | 1898 | 0 | 0 | 266 | 40 | 7620 | 10352 |
| 2 | 654 | 6156 | 11068 | 3136 | 174 | 44 | 11494 | 232 | 50262 | 83220 |
| 3 | 28 | 0 | 3894 | 426 | 276 | 30 | 1504 | 164 | 22162 | 28484 |
| 4 | 0 | 21850 | 41252 | 15740 | 4404 | 0 | 1086 | 0 | 400000 | 484332 |
| 5 | 0 | 0 | 1110 | 17500 | 46 | 0 | 596 | 0 | 85356 | 104608 |
| 6 | 0 | 9196 | 386 | 2752 | 430 | 67528 | 218 | 0 | 28318 | 108828 |
| 7 | 108 | 0 | 4344 | 12590 | 36 | 0 | 3366 | 114 | 52550 | 73108 |
| 8 | 0 | 0 | 63104 | 3628 | 0 | 0 | 214 | 44 | 17404 | 84394 |
| 9 | 124 | 3164 | 68764 | 7248 | 10902 | 2922 | 9950 | 1384 | 200000 | 304458 |
| 10 | 0 | 0 | 2240 | 3904 | 3084 | 7698 | 1296 | 54 | 20050 | 38326 |
| 11 | 0 | 0 | 5350 | 4594 | 3162 | 1208 | 616 | 0 | 26284 | 41214 |
| 12 | 1630 | 540 | 7570 | 31616 | 2524 | 41010 | 7034 | 232 | 105336 | 197492 |
| 13 | 86 | 4976 | 16700 | 13454 | 0 | 0 | 1192 | 30 | 15050 | 51488 |
| 14 | 146 | 8900 | 1440 | 122 | 574 | 9058 | 1736 | 30 | 13346 | 35352 |
| 15 | 0 | 12416 | 43894 | 2200 | 46 | 0 | 9270 | 798 | 79914 | 148538 |
| 16 | 0 | 6134 | 662 | 860 | 62 | 0 | 106 | 228 | 1206 | 9258 |
| 17 | 0 | 0 | 150 | 992 | 0 | 0 | 38 | 168 | 1238 | 2586 |
| 18 | 278 | 4156 | 1672 | 16908 | 0 | 0 | 784 | 100 | 1622 | 25520 |
| 19 | 316 | 3978 | 3310 | 4892 | 25412 | 19282 | 636 | 214 | 23902 | 81942 |
| 20 | 0 | 170 | 22852 | 6626 | 32 | 0 | 1716 | 814 | 88798 | 121008 |
| 21 | 94 | 7654 | 242 | 3442 | 342 | 114 | 398 | 0 | 17390 | 29676 |
| 22 | 0 | 0 | 444 | 850 | 0 | 0 | 144 | 0 | 3876 | 5314 |
| 23 | 0 | 0 | 51642 | 1294 | 56 | 0 | 406 | 204 | 14712 | 68314 |
| 24 | 320 | 2204 | 1922 | 5006 | 58032 | 19986 | 768 | 674 | 16020 | 104932 |
| 25 | 0 | 358 | 1852 | 3962 | 0 | 0 | 322 | 0 | 11892 | 18386 |
| 26 | 0 | 0 | 24872 | 1320 | 0 | 0 | 322 | 42 | 12594 | 39150 |
| 27 | 28 | 6658 | 5460 | 27430 | 13330 | 9800 | 15910 | 68 | 40730 | 119414 |

For the experiments, both individual and fusion classifiers employed the FF-MLP Neural Network with a Levenberg-Marquardt backpropagation learning algorithm, 100 epochs, a Tan-Sigmoid transfer function, and one hidden layer, with the neuron count varying between 10 and 30. The dataset was evenly split into training and test sets to ensure unbiased performance evaluation. Each individual 2-class classifier was trained with one user serving as the authorised user, while all other users acted as impostors, following a standard testing strategy. A threshold of 28 interactions per application per user was set to ensure a minimum sample size for classification. After training, there were a total of 27 individual 2-class classifiers for the second and third experiments.

## 4 Experimental Results

The results of experiment one multi-classifier identification approach yielded average True Positive Identification Rate (TPIR) of 50.2%, 64.5% and 71% for ranks 1, 3 and 5, respectively, as shown in Table **5**. In comparison, the single classifier approach by (Clarke et al. 2017) resulted in error rates of 47.5%, 60.5%, and 66% for the corresponding ranks. Analysing individual performances, the highest rank 1 performance was achieved by participant 27 with a TPIR of 88.5%. This performance increased to 92.8% within rank 5. Conversely, the lowest-performing participant was participant number 22, with a rank 1 performance of 19%, which improved to 68.6% by rank 5. The primary goal of this identification process is to prioritise traffic and reduce data volume for investigators, making rank 1 identification non-essential. These results highlight significant capabilities in achieving this objective.

**Table 5: TPIR Ranks and Fusion Results for Users**

| User ID | TPIR Rank 1 (%) | | TPIR Rank 3 (%) | | TPIR Rank 5 (%) | |
|---|---|---|---|---|---|---|
| | Fusion | Multi-Classifier | Fusion | Multi-Classifier | Fusion | Multi-Classifier |
| 1 | 50.5 | 49 | 59 | 55.3 | 66.9 | 63.9 |
| 2 | 64.4 | 48.2 | 73.9 | 70.1 | 79.8 | 74.4 |
| 3 | 55.4 | 46.3 | 68.3 | 64.8 | 81.2 | 74.4 |
| 4 | 80.5 | 65.8 | 89 | 78.3 | 90.7 | 82.1 |
| 5 | 45.9 | 32.3 | 64.3 | 38 | 69 | 51.7 |
| 6 | 61 | 55.1 | 71.9 | 71.8 | 77.9 | 79.8 |
| 7 | 47.6 | 36.9 | 60.3 | 69.1 | 70.2 | 80.2 |
| 8 | 70.5 | 60.8 | 75.9 | 67.6 | 78.9 | 68.7 |
| 9 | 82 | 68.6 | 89.4 | 75.4 | 92.8 | 82.6 |
| 10 | 51.7 | 39.4 | 69 | 56.9 | 74.6 | 63.9 |
| 11 | 65.4 | 51.2 | 78.1 | 55.1 | 84.2 | 57.9 |
| 12 | 72.3 | 65.3 | 82.4 | 75 | 85.4 | 78.8 |
| 13 | 60.1 | 54.1 | 71.2 | 63.3 | 82 | 69.5 |
| 14 | 50.7 | 34.9 | 64.6 | 62.5 | 69 | 72.7 |
| 15 | 67.8 | 59.8 | 85.3 | 80.1 | 88.2 | 84.9 |
| 16 | 40.5 | 31.1 | 65.5 | 53.2 | 71.4 | 60.1 |
| 17 | 31.4 | 28.4 | 43.9 | 39.1 | 46.3 | 43.6 |
| 18 | 66.9 | 64.1 | 80.3 | 73.6 | 89.7 | 75.7 |
| 19 | 55.7 | 45.5 | 71.5 | 60.1 | 77 | 71.2 |
| 20 | 64.1 | 44.7 | 74.5 | 51.5 | 86.4 | 64.2 |
| 21 | 50 | 50.6 | 69.1 | 71.8 | 79.3 | 87 |
| 22 | 36.3 | 19.1 | 72.3 | 66.4 | 80.9 | 68.6 |
| 23 | 55.7 | 41 | 70.5 | 54.4 | 73.3 | 61.7 |
| 24 | 79.2 | 79.5 | 87.3 | 84.2 | 94.1 | 85.8 |
| 25 | 47.2 | 53.7 | 55.1 | 61.2 | 60.2 | 68.3 |
| 26 | 63.2 | 50.2 | 68.6 | 52.9 | 72.4 | 54.8 |
| 27 | 90 | 88.5 | 95.6 | 91.6 | 98.6 | 92.8 |
| **Average** | 59.4 | 50.2 | 72.4 | 64.5 | 78.5 | 71 |

Table **5** also reveals fusion-based results (experiment two), demonstrating an average 10-18% TPIR

improvement compared to selecting the highest output value. Most users experienced enhanced TPIR performance across ranks 1, 3, and 5. When evaluating recognition performance by services, Skype and Hotmail exhibited strong discrimination abilities. Utilising the Fusion approach, TPIR exceeded 70% for all services except Dropbox (Table 6). While participant numbers varied across services, no significant relationship emerged in error rates among different participant groups.

**Table 6: Recognition Performance based upon Service**

| Application Name | Number of Users | Rank 1 (%) | | Rank 3 (%) | | Rank 5 (%) | |
|---|---|---|---|---|---|---|---|
| | | Fusion | Standard | Fusion | Standard | Fusion | Standard |
| Skype | 12 | 99.8 | 98.1 | 100 | 98.2 | 100 | 98.2 |
| Hotmail | 19 | 97.3 | 96.2 | 98.9 | 96.9 | 99.3 | 97 |
| Facebook | 27 | 83.4 | 66.7 | 87.6 | 70.8 | 88.6 | 71.9 |
| BBC | 12 | 83.1 | 81.8 | 93.2 | 92.5 | 97 | 95.4 |
| Google | 27 | 82.1 | 71.7 | 88.8 | 79.4 | 90.4 | 82.2 |
| Wikipedia | 20 | 72.7 | 66.9 | 85.8 | 83.6 | 90.3 | 89.2 |
| Twitter | 27 | 72.4 | 65.3 | 85 | 79.5 | 89.3 | 83.4 |
| YouTube | 27 | 71.4 | 62.8 | 84 | 74.8 | 87.9 | 78.9 |
| Dropbox | 16 | 66.6 | 57.1 | 79.3 | 73.9 | 84.5 | 82.8 |

The primary objective of the algorithm is to establish a 'proof of life' and capture the temporary IP addresses in use. Table 7 demonstrates the practical performance, showing that an individual's best service network traffic can be identified in 93% of cases on average using fusion. Even the second and third services can be successfully classified in 83% and 69% of cases, respectively, serving as strong 'proof of life' indicators.

**Table 7: Best Service Recognition Performance**

| User ID | First Application | | | | Second Application | | | | Third Application | | | |
|---|---|---|---|---|---|---|---|---|---|---|---|---|
| | Fusion | | Multi-classifier | | Fusion | | Multi-classifier | | Fusion | | Multi-classifier | |
| | Name | TPIR (%) | Name | TPIR (%) | Name | TPIR (%) | Name | TPIR (%) | Name | TPIR (%) | Name | TPIR (%) |
| 1 | Wiki. | 100 | Wiki. | 100 | Google | 95.2 | Google | 95.2 | YouTube | 57.8 | YouTube | 50 |
| 2 | Skype | 94.1 | Skype | 94.1 | BBC | 75.5 | BBC | 74.6 | Wiki. | 74.1 | Wiki. | 74.1 |
| 3 | Skype | 100 | Skype | 100 | Google | 75.1 | Hotmail | 60.8 | Twitter | 74.6 | Google | 59.1 |
| 4 | Google | 93.5 | Hotmail | 91.8 | Hotmail | 91.9 | Google | 91.6 | YouTube | 89.9 | YouTube | 90.9 |
| 5 | YouTube | 82.2 | YouTube | 74.1 | Google | 80.1 | Google | 73.6 | Hotmail | 40.2 | Hotmail | 13 |
| 6 | Skype | 100 | Skype | 100 | Hotmail | 92 | Google | 89.1 | Google | 88.3 | Hotmail | 85.5 |
| 7 | Google | 86.7 | Google | 79.2 | BBC | 81.4 | BBC | 64.8 | Wiki. | 50 | Wiki. | 50 |
| 8 | Wiki. | 100 | Wiki. | 100 | Facebook | 90.6 | Facebook | 79.5 | YouTube | 61.3 | YouTube | 56.5 |
| 9 | Skype | 100 | Skype | 100 | Hotmail | 96.4 | Hotmail | 95 | Wiki. | 93.2 | Wiki. | 93 |
| 10 | Skype | 100 | Hotmail | 83 | Hotmail | 86.4 | Skype | 63.7 | Google | 65.7 | Google | 56.1 |
| 11 | Hotmail | 95 | Hotmail | 80.5 | Skype | 85.4 | Skype | 80.3 | Facebook | 71.1 | Google | 72.7 |
| 12 | Skype | 100 | Skype | 99.7 | BBC | 97.4 | BBC | 95 | Google | 85.8 | Hotmail | 80.2 |
| 13 | Dropbox | 89.5 | Dropbox | 80.9 | Facebook | 88.1 | Google | 75.3 | Google | 81.2 | Facebook | 72.7 |
| 14 | Skype | 100 | Skype | 100 | Hotmail | 72.4 | Hotmail | 62.3 | Twitter | 63 | Dropbox | 48.2 |
| 15 | Facebook | 89.3 | Facebook | 74.5 | Dropbox | 82.5 | YouTube | 71.1 | YouTube | 76.9 | Dropbox | 70.9 |

| 16 | Wiki. | 95.6 | Wiki. | 95.6 | Google | 71.1 | Hotmail | 35.4 | YouTube | 43.6 | YouTube | 28 |
|---|---|---|---|---|---|---|---|---|---|---|---|---|
| 17 | Google | 68.3 | Google | 59 | Wiki. | 57.1 | Wiki. | 52 | YouTube | 31.6 | YouTube | 30.6 |
| 18 | Wiki. | 98 | Wiki. | 98 | BBC | 86.3 | BBC | 82 | Google | 76.7 | YouTube | 67.2 |
| 19 | Skype | 99.4 | Skype | 99.4 | Hotmail | 97.5 | Hotmail | 95 | BBC | 70.8 | BBC | 61.7 |
| 20 | Dropbox | 100 | Dropbox | 75.2 | Facebook | 83.5 | Facebook | 73.9 | Google | 80.9 | Google | 63.7 |
| 21 | Skype | 100 | Skype | 100 | Twitter | 78.3 | Hotmail | 85.3 | BBC | 72.3 | Twitter | 79.3 |
| 22 | Twitter | 65.2 | Twitter | 43 | Google | 41.1 | Google | 29.4 | YouTube | 38.9 | YouTube | 4.2 |
| 23 | Facebook | 92.5 | Facebook | 75.5 | Hotmail | 75 | Hotmail | 71.4 | Twitter | 58.6 | Twitter | 58.6 |
| 24 | Hotmail | 100 | Hotmail | 100 | Skype | 100 | Skype | 100 | BBC | 84.5 | BBC | 91.8 |
| 25 | Google | 85.4 | Google | 80.8 | Dropbox | 75.9 | Dropbox | 75.9 | YouTube | 58.3 | YouTube | 51.1 |
| 26 | Facebook | 85.6 | Wiki. | 76.1 | Google | 72.4 | Google | 64.8 | Wiki. | 71.4 | Facebook | 62.4 |
| 27 | Skype | 100 | Skype | 100 | Google | 100 | Google | 100 | YouTube | 100 | YouTube | 100 |
| **Avg.** | | 93.3 | | 87.4 | | 82.5 | | 75 | | 68.9 | | 61.9 |

The final analysis examined the impact of utilising temporary IP addresses to group data sent within defined time windows before and after them. Using the outcomes of the fusion approach and rank 1 recognition, time windows of 30 seconds, 60 seconds, and 240 seconds were sequentially tested. These time windows were chosen to minimise the likelihood of IP reallocation, especially due to mobile devices being powered off. As displayed in Table 8, employing timeline analysis increased the average performance from 59% to 70% with a 30-second time window. Although performance continued to improve with larger time windows, reaching 73% with a 240-second window, the potential for IP reassignment and the limited performance gain beyond 30 seconds suggest that a 30-second time window offers the optimal trade-off between performance and IP reassignment.

**Table 8: User Recognition Performance using Timeline Analysis**

| User ID | Fusuib (%) | Timeline Analysis (%) | | |
|---|---|---|---|---|
| | TPIR Rank 1 | 30 Seconds | 60 Seconds | 240 Seconds |
| 1 | 50.5 | 68.6 | 69.6 | 71 |
| 2 | 64.4 | 46.6 | 47.2 | 50.5 |
| 3 | 55.4 | 55.5 | 56.6 | 58.1 |
| 4 | 80.5 | 93.1 | 94 | 96.7 |
| 5 | 45.9 | 88 | 88.7 | 89.5 |
| 6 | 61 | 79.8 | 79.8 | 79.8 |
| 7 | 47.6 | 54.9 | 54.9 | 55.2 |
| 8 | 70.5 | 91.3 | 92.1 | 93.6 |
| 9 | 82 | 86.5 | 86.9 | 88 |
| 10 | 51.7 | 66 | 67.4 | 72.1 |
| 11 | 65.4 | 61.1 | 62.7 | 67.7 |
| 12 | 72.3 | 77.7 | 80.1 | 81.8 |

| 13 | 60.1 | 81.3 | 83.1 | 84.6 |
|---|---|---|---|---|
| 14 | 50.7 | 64.7 | 65.4 | 72.9 |
| 15 | 67.8 | 90.4 | 91.7 | 93.2 |
| 16 | 40.5 | 46 | 46.1 | 46.8 |
| 17 | 31.4 | 48.6 | 48.6 | 49.8 |
| 18 | 66.9 | 72.5 | 74.3 | 76.9 |
| 19 | 55.7 | 76.1 | 77.2 | 79.1 |
| 20 | 64.1 | 67.1 | 67.4 | 67.7 |
| 21 | 50 | 26.1 | 26.1 | 26.4 |
| 22 | 36.3 | 37.6 | 38.2 | 38.5 |
| 23 | 55.7 | 84.1 | 85.5 | 86.5 |
| 24 | 79.2 | 94.8 | 94.8 | 94.8 |
| 25 | 47.2 | 61.3 | 62.6 | 64.6 |
| 26 | 63.2 | 79.5 | 81.1 | 83.2 |
| 27 | 90 | 97.5 | 97.6 | 97.7 |
| Average | 59.4 | 70.2 | 71.1 | 72.8 |

The evaluation results highlight the distinctiveness of user interactions, offering a reliable means of user identification. Implementing a time window in this method would lead to even higher recognition performance. Across all service usage, many participants achieve recognition performance that significantly reduces the network data load for forensic investigators. With suitable interfaces facilitating in-depth exploration of interactions and raw data, this approach promises substantial time and cognitive load reduction.

## 5 Discussion

This research serves as the basis for proposing an innovative user-focused N-FAT. It stems from an analysis that identified essential requirements. These include: extracting valuable insights from encrypted network data to understand user activities within network-based applications, analysing traffic from a user-centric viewpoint rather than just an IP address, offering analysis flexibility from packets to interactions, and providing forensic tools for higher-level data queries. Additionally, data visualisation aids in more reliable data interpretation. These requirements collectively improve evidence identification from vast low-level data, reduce investigator cognitive load in recognising relationships among artefacts, and subsequently lower investigation time and costs. In line with established FS-FAT principles and acknowledging the rising need for collaborative investigations, we identified extra requirements, including comprehensive case management, robust authentication and authorisation, platform-agnostic tools, centralised resources, and multi-user capabilities. Utilising a web application via a private network enables investigators to access and process cases using standard web browsers, reducing workstation computational demands and shifting tasks to a scalable cloud-based infrastructure. This approach would provide a more user-friendly, flexible, and cost-effective alternative to traditional infrastructure methods.

This visual representation offers insights into applications used by different users, facilitating the rapid detection of unusual usage patterns within an organisation. Figure 2 displays a timeline analysis focused on a single user's application usage. While user and interaction identification may occasionally produce incorrect classifications, their application in the N-FAT system aims to reduce data volume and prioritise investigator queries. For instance, a query about a specific user activity at a certain time can lead to further investigation by pivoting on the resulting IP address. The flexibility of the visualisations allows for the integration of additional tools into the system. For each query that is performed against the data, the investigator would have the opportunity to bookmark the results should they wish to. This is managed by the Reporting function, and an example is illustrated in Figure 3. These bookmarks contain the visualisation, a comments section for freeform notes to be added by the investigator, the database queries utilised to generate the visualisation and the filtering options applied to the data. The bookmarked data also includes the extracted raw data that the query is based upon. The raw data is provided because this is the true source data that can be relied upon. The interactions and identification of users are subject to error to confirmation of what is being seen in the visualisation is provided for examination by all parties.

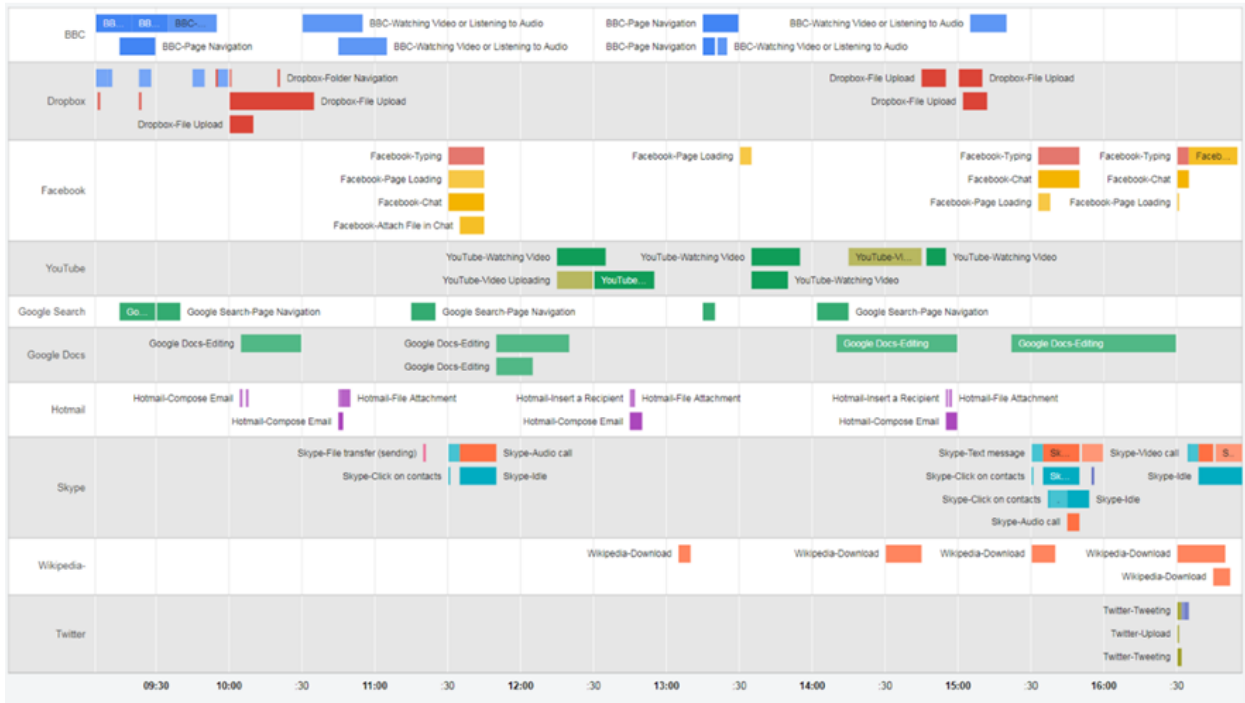

**Figure 2: Timeline Analysis of User Interactions**

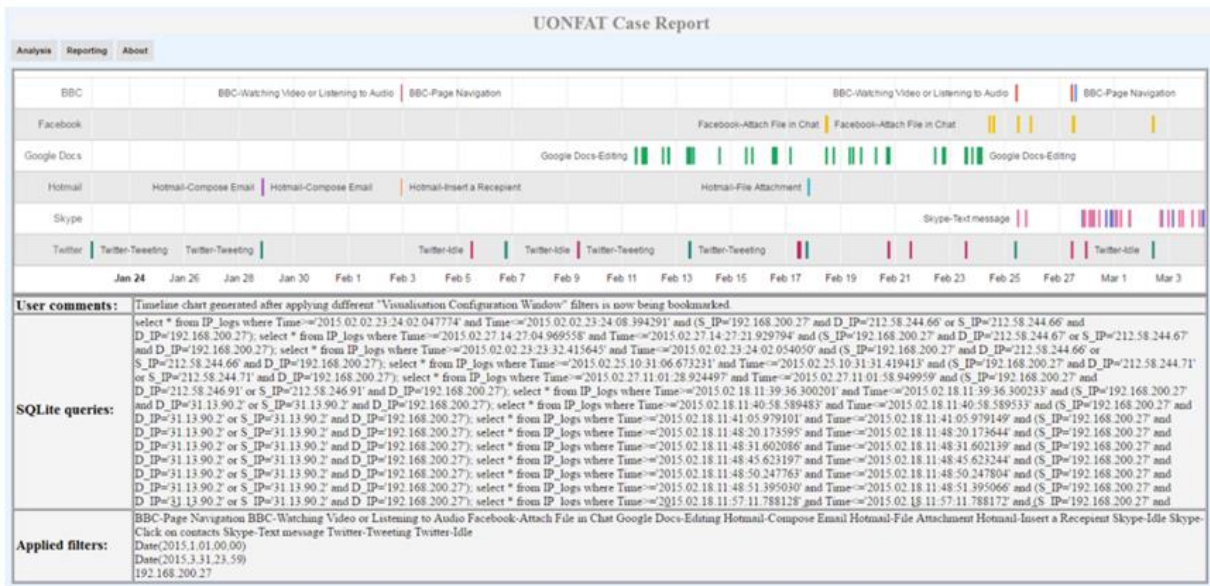

**Figure 3: Extract of the Case Report Entry**

In comparison to existing N-FATs, this approach offers a graphical user interface that explores data based upon interactions and users rather than low-level IP addresses. The visualisations themselves provide the ability to interact with data objects in a flexible manner whilst continuously maintaining a mapping back to the raw traffic. This should help identify relevant relationships and do so in a timely fashion. The ability to perform this analysis on completely encrypted traffic sets itself ahead of the current state of the art.

## 6 Conclusion

This study has introduced an innovative N-FAT dedicated to analysing and investigating user interactions with network-based services. The presented approach is robust, flexible, and extensible that demonstrates abstracted network data in a more usable and cognitively manageable visualisation. Additional research is needed to automate the identification of new interactions, potentially using a hybrid deterministic and probabilistic approach. Furthermore, research should explore the nature of biometric templates used for user identification, particularly regarding their permanence and update frequency to reflect current user behaviour. Finally, a comprehensive system evaluation by stakeholders is imperative, with particular attention to interface design and the potential incorporation of supplementary functionality and forensic analyses.

## References

Abuadlla, Y., Kvascev, G., Gajin, S., and Jovanovic, Z. (2014). Flow-based anomaly intrusion detection system using two neural network stages. *Computer Science and Information Systems*, *11*(2), 601–622. https://doi.org/10.2298/CSIS130415035A

Ahmed, I., and Lhee, K. (2011). Classification of packet contents for malware detection. *Journal in Computer Virology*, *7*(4), 279–295. https://doi.org/10.1007/s11416-011-0156-6

Al-Bataineh, A., and White, G. (2012). Analysis and detection of malicious data exfiltration in web traffic. *2012 7th International Conference on Malicious and Unwanted Software*, 26–31. https://doi.org/10.1109/MALWARE.2012.6461004

Alotibi, G., Clarke, N., Fudong Li, and Furnell, S. (2016). User profiling from network traffic via novel application-level interactions. *2016 11th International Conference for Internet Technology and Secured Transactions (ICITST)*, 279–285. https://doi.org/10.1109/ICITST.2016.7856712

Boukhtouta, A., Mokhov, S. A., Lakhdari, N.-E., Debbabi, M., and Paquet, J. (2016). Network malware classification comparison using DPI and flow packet headers. *Journal of Computer Virology and Hacking Techniques*, *12*(2), 69–100. https://doi.org/10.1007/s11416-015-0247-x

Clarke, N., Li, F., and Furnell, S. (2017). A novel privacy preserving user identification approach for network traffic. *Computers and Security*, *70*, 335–350. https://doi.org/10.1016/j.cose.2017.06.012

Fernandes, G., Rodrigues, J. J. P. C., and Proença, M. L. (2015). Autonomous profile-based anomaly detection system using principal component analysis and flow analysis. *Applied Soft Computing*, *34*, 513–525. https://doi.org/10.1016/j.asoc.2015.05.019

He, G., Zhang, T., Ma, Y., and Xu, B. (2014). A Novel Method to Detect Encrypted Data Exfiltration. *2014 Second International Conference on Advanced Cloud and Big Data*, 240–246. https://doi.org/10.1109/CBD.2014.40

Hofstede, R., Bartos, V., Sperotto, A., and Pras, A. (2013). Towards real-time intrusion detection for NetFlow and IPFIX. *Proceedings of the 9th International Conference on Network and Service Management (CNSM 2013)*, 227–234. https://doi.org/10.1109/CNSM.2013.6727841

Khan, S., Gani, A., Wahab, A. W. A., Shiraz, M., and Ahmad, I. (2016). Network forensics:

Review, taxonomy, and open challenges. *Journal of Network and Computer Applications*, *66*, 214–235. https://doi.org/10.1016/j.jnca.2016.03.005

Leroux, S., Bohez, S., Maenhaut, P.-J., Meheus, N., Simoens, P., and Dhoedt, B. (2018). Fingerprinting encrypted network traffic types using machine learning. *NOMS 2018 - 2018 IEEE/IFIP Network Operations and Management Symposium*, 1–5. https://doi.org/10.1109/NOMS.2018.8406218

Meghdouri, F., Vazquez, F. I., and Zseby, T. (2020). Cross-Layer Profiling of Encrypted Network Data for Anomaly Detection. *2020 IEEE 7th International Conference on Data Science and Advanced Analytics (DSAA)*, 469–478. https://doi.org/10.1109/DSAA49011.2020.00061

Parvat, T. J., and Chandra, P. (2015). A Novel Approach to Deep Packet Inspection for Intrusion Detection. *Procedia Computer Science*, *45*(C), 506–513. https://doi.org/10.1016/j.procs.2015.03.091

Saevanee, H., Clarke, N., Furnell, S., and Biscione, V. (2015). Continuous user authentication using multi-modal biometrics. *Computers and Security*, *53*, 234–246. https://doi.org/10.1016/j.cose.2015.06.001

Shafiee Hasanabadi, S., Habibi Lashkari, A., and Ghorbani, A. A. (2020). A survey and research challenges of anti-forensics: Evaluation of game-theoretic models in simulation of forensic agents' behaviour. *Forensic Science International: Digital Investigation*, *35*. https://doi.org/10.1016/j.fsidi.2020.301024

Stergiopoulos, G., Talavari, A., Bitsikas, E., and Gritzalis, D. (2018). Automatic Detection of Various Malicious Traffic Using Side Channel Features on TCP Packets. In *Lecture Notes in Computer Science (including subseries Lecture Notes in Artificial Intelligence and Lecture Notes in Bioinformatics): Vol. 11098 LNCS* (pp. 346–362). Springer International Publishing. https://doi.org/10.1007/978-3-319-99073-6_17

Stevanovic, M., and Pedersen, J. M. (2014). An efficient flow-based botnet detection using supervised machine learning. *2014 International Conference on Computing, Networking and Communications, ICNC 2014*, 797–801. https://doi.org/10.1109/ICCNC.2014.6785439

Taylor, V. F., Spolaor, R., Conti, M., and Martinovic, I. (2016). AppScanner: Automatic Fingerprinting of Smartphone Apps from Encrypted Network Traffic. *2016 IEEE European Symposium on Security and Privacy (EuroSandP)*, 439–454. https://doi.org/10.1109/EuroSP.2016.40

Tcpdump. (2023). *Tcpdump manual page*. 2023. https://www.tcpdump.org/manpages/tcpdump.1.html

Tegeler, F., Fu, X., Vigna, G., and Kruegel, C. (2012). BotFinder. *Proceedings of the 8th International Conference on Emerging Networking Experiments and Technologies*, 349–360. https://doi.org/10.1145/2413176.2413217

Wireshark. (2023, December). *About wireshark*. 2023. https://www.wireshark.org/about.html